\documentclass[journal=jacsat,manuscript=article]{achemso}
\usepackage{setspace}
\usepackage{amsmath,amssymb}
\usepackage{graphicx}
\usepackage{dcolumn}
\usepackage{bm}
\usepackage{multirow}
\usepackage{natbib}
\setkeys{acs}{articletitle = true}

\title{Generalized Kasha's Scheme for Classifying Two-Dimensional Excitonic Molecular Aggregates: Temperature Dependent Absorption Peak Frequency Shift}
\author{Chern Chuang}
\affiliation{Department of Chemistry, Massachusetts Institute of Technology, MA 02139, USA}
\author{Doran I. G. Bennett}
\affiliation{Department of Chemistry and Chemical Biology, Harvard University, Cambridge, MA 02138, USA}
\alsoaffiliation{Bio-Inspired Solar Energy Program, Canadian Institute for Advanced Research, Toronto, ON M5G 1Z8, Canada}
\author{Justin R. Caram}
\affiliation{Department of Chemistry and Biochemistry, University of California, Los Angeles, CA 90095, USA}
\author{Al\'{a}n Aspuru-Guzik}
\affiliation{Department of Chemistry and Chemical Biology, Harvard University, Cambridge, MA 02138, USA}
\alsoaffiliation{Bio-Inspired Solar Energy Program, Canadian Institute for Advanced Research, Toronto, ON M5G 1Z8, Canada}
\alsoaffiliation{Department of Chemistry and Department of Computer Science, University of Toronto, Toronto, Ontario M5S 3H6, Canada and Vector Institute, Toronto, ON M5G 1M1, Canada}
\author{Moungi G. Bawendi}
\affiliation{Department of Chemistry, Massachusetts Institute of Technology, MA 02139, USA}
\author{Jianshu Cao}
\affiliation{Department of Chemistry, Massachusetts Institute of Technology, MA 02139, USA}
\email{jianshu@mit.edu}

\begin{document}

\begin{abstract}
We propose a generalized theoretical framework for classifying two-dimensional (2D) excitonic molecular aggregates based on an analysis of temperature dependent spectra. In addition to the monomer-aggregate absorption peak shift, which defines the conventional J- and H-aggregates, we incorporate the peak shift associated with increasing temperature as a measure to characterize the exciton band structure. First we show that there is a one-to-one correspondence between the monomer-aggregate and the temperature dependent peak shifts for Kasha's well-established model of one-dimensional (1D) aggregates, where J-aggregates exhibit further redshift upon increasing temperature and H-aggregates exhibit further blueshift. On the contrary, 2D aggregate structures are capable of supporting the two other combinations: blueshifting J-aggregates and redshifting H-aggregates, owing to their more complex exciton band structures. Secondly, using standard spectral lineshape theory, the temperature dependent shift is associated with the relative abundance of states on each side of the bright state. We further establish that the density of states can be connected to the microscopic packing condition leading to these four classes of aggregates by separately considering the short and long-range contribution to the excitonic couplings. In particular the temperature dependent shift is shown to be an unambiguous signature for the sign of net short-range couplings: Aggregates with net negative (positive) short-range couplings redshift (blueshift) with increasing temperature. Lastly, comparison with experiments shows that our theory can be utilized to quantitatively account for the observed but previously unexplained $T$-dependent absorption lineshapes. Thus, our work provides a firm ground for elucidating the structure-function relationships for molecular aggregates and is fully compatible with existing experimental and theoretical structure characterization tools.
\end{abstract}

\maketitle
\section{Introduction}
Organic optoelectronic materials have been in the spotlight of energy research for decades.\cite{Davydov,KenkreReineker,Mobius1999,Wurthner2011,Bardeen2014,SpanoACR2017} One example is the ubiquitous pigment-protein complexes occurring in light-harvesting bacteria and plants. Their efficiency in transforming the solar energy into biomass and robustness at adapting to vastly different and rapidly changing environments can provide insight into the design of new functional materials.\cite{Cogdell2008,Blankenship2011} Excitonic molecular aggregates are one promising candidate. 
Instead of a protein scaffold that supports and confines the light-absorbing pigments, a molecular aggregate is self-assembled from its constituent pigment molecules owing to their intermolecular interactions (including polarizability, hydrophobic and hydrophilic functional groups). The resulting aggregates can have superstructures with a variety of geometries and topologies.\cite{Kirstein2006}

In the 1930s Jelly\cite{Jelly1936} and Scheibe\cite{Scheibe1937} independently discovered excitonic aggregates and associated the observed spectral redshift and narrowing to the interactions among the constituent dyes. The diversity of chromophores combined with the possible aggregate superstructures provides tremendous tunability in spectroscopic properties.\cite{SpanoACR2010,TubePRL} The exact connection between the spectroscopic properties of monomers and those of aggregates, however, have not been established. While the spectral narrowing indicates a high degree of spatial homogeneity across the aggregates, straightforward inference of the packing conditions from spectroscopies is not possible. Indirect methods combining spectroscopic information with theoretical structural modeling are generally necessary. \cite{JPCB.101.5646,Wurthner2011,EisfeldJCP2013,Wurthner2017,EisfeldPRL2017} We should note that modern structural characterization tools such as cryo-EM can only provide limited microscopic information on the packing conditions, which often depend subtly on their solvation environment, not captured in cryo-EM. \cite{deGroot2009,Friedl2016} This is in sharp contrast to molecular crystals or the previously mentioned pigment-protein complexes where x-ray diffraction offers atomistic details. 

An important step toward establishing such structure-function relationship in molecular aggregates was taken by Kasha.\cite{Kasha1963,Kasha1965}. In Kasha's theory, the influence of the aggregate geometry on the emergent spectroscopic properties can be summarized by a single parameter $\alpha$, the angle of the molecular transition dipole makes with the axis of aggregation. He found that with a small $\alpha$, a head-to-tail configuration, leads to negative excitonic couplings and a redshifted system bright state, a J-aggregate. On the other hand, a larger $\alpha$ corresponding to a side-by-side configuration leads to positive couplings and a blueshifted bright state, an H-aggregate. A large monomer-aggregate absorption peak shift indicates strong intermolecular excitonic couping, and the exciton center-of-mass wavefunction delocalizes over multiple molecules. This further gives rise to a reduction of inhomogeneous broadening, referred to as the motional narrowing. \cite{Knapp1984,Malyshev1993,Walczak2008}

However, Kasha's theory is less useful at inferring the packing conditions from linear spectra for aggregate superstructures more complicated than a dimer or simple 1D chains. \cite{Nabetani1995,Yamaguchi2006,Spano2012,Dylan2013,ChemRev2014,Arend2D,SpanoACR2017,SpanoCR2017} For example, it is found that in the 5,5',6,6'-tetrachlorobenzimidacarbocyanine dyes of various hydrophilic and hydrophobic side chain lengths, the superstructure varies from linear (1D) to planar (2D), tubular (quasi-1D), or even bundles of tubes (ribbons).\cite{Kirstein2006} With the exception of the case where all transition dipoles aligned perpendicular to the aggregate plane, the inherent incommensurability between dipole-dipole interactions among the constituent dyes and the 2D topology forbids a binary assignment of specific geometric parameters from the monomer-aggregate peak shift.\cite{Czikkely1970,Turlet1992,KuhnMobius1993,Witten1994,Suarez2003,Tichler2015} Extensive numerical simulations and fittings, of presumed geometric models with tunable parameters, are generally required to assess the packing conditions.\cite{JPCA.114.8179}

In this contribution we propose a generalized classification scheme similar to Kasha's theory for 1D aggregates and demonstrate its applicability to 2D aggregates for extracting key microscopic packing parameters. This is possible using temperature-dependent linear spectroscopy and taking into account the direction of absorption peak shift with increasing temperature, which is easily available experimentally owing to the establishment of sucrose encapsulation methods.\cite{JustinNanoLett2016,ACSNano2018} Using standard lineshape theory we show that in 2D models the $T$-dependent peak shift reveals additional information about its excitonic DoS in the vicinity of the bright state, leading to the existence of blueshifting J-aggregates and redshifting H-aggregates as previously observed experimentally without explanation.\cite{Yamaguchi2006,Wurthner2009} This is in contrast to 1D models where the absorption peaks of J-aggregates always redshift upon increasing temperature, while the opposite is true for H-aggregates. We establish the connection between the direction of the $T$-dependent shift and the sign of the total short-range couplings by separately considering the short and the long-range coupling contributions to the excitonic DoS. Specifically we show that an aggregate with redshifting absorption peak upon increasing temperature corresponds to one with a net negative short-range coupling. In addition, we show that the long-range coupling is completely determined by a single parameter, the zenith angle of the transition dipole moment with respect to the plane of aggregation. The combination of these knowledge provides a wealth of unambiguous information on the packing conditions and paves the way to comprehensively understand the structure-function relationships in these systems.

We start by reviewing the necessary lineshape theory for interpreting $T$-dependent absorption spectra of excitonic aggregates. This is first applied to the traditional 1D models in Kasha's scheme and then to 2D aggregates, where we predict two new types of excitonic aggregates. This is then related to the relative position of their bright states to the excitonic DoS. We identify the zenith angle of the transition dipole moment as a key parameter based on an analysis separating the contributions of the short- and the long-range couplings, and introduce a 2D configurational diagram to help facilitate parameter space exploration. This naturally leads to a connection to the packing conditions as we demonstrate that the short-range component dominates in determining the $T$-dependent peak shift. We also apply the theoretical framework to a real system, a family of merocyanine dye monolayers. We demonstrate quantitative agreement between theory and experiment and extract essential microscopic packing information.

\section{Temperature Dependent Absorption Lineshape of Delocalized Frenkel Excitons}
In this section we review the spectroscopic lineshape theory necessary to explain the temperature dependent absorption spectra of excitonic aggregates. Starting from the standard Frenkel exciton Hamiltonian
\begin{eqnarray}
\hat{H}_\mathrm{s}=\sum_{\vec{n}}\epsilon_{\vec{n}}|{\vec{n}}\rangle\langle {\vec{n}}|+\sum_{{\vec{n}}\ne{\vec{m}}}J_{{\vec{n}},{\vec{m}}}|{\vec{n}}\rangle\langle{\vec{m}}|,
\label{eqn:Hs}
\end{eqnarray}
where $\epsilon_{\vec{n}}$ and $J_{{\vec{n}},{\vec{m}}}$ are the site energy and the excitonic coupling, $|{\vec{n}}\rangle$ represents the state where the designated site is in the excited state while all other sites in the ground state. We consider the homogeneous limit, where $\epsilon_{\vec{n}}$ are set to the same value independent of $\vec{n}$ and $J_{{\vec{n}},{\vec{m}}}=J_{{\vec{n}}-{\vec{m}}}$ is translationally invariant.\cite{JianFCE} We set $\hbar$ to unity throughout this paper for brevity. To introduce thermal effects, the environmental degrees of freedom are represented as a collection of harmonic oscillators coupled linearly and independently to each molecule
\begin{eqnarray}
\hat{H}_\mathrm{b}=\sum_i\omega_i\hat{b}_i^\dagger\hat{b}_i,\hat{H}_\mathrm{sb}=\sum_{\vec{n}}\sum_ig_{{\vec{n}},i}\left(\hat{b}_i^\dagger+\hat{b}_i\right)|{\vec{n}}\rangle\langle{\vec{n}}|
\end{eqnarray}
where $\hat{b}_i^\dagger$ and $\hat{b}_i$ creates and annihilates a phonon with frequency $\omega_i$ and $g_{{\vec{n}},i}$ is the exciton-phonon coupling strength. Typically one expresses the set of bath modes by a continuous profile, for example an Ohmic spectral density $J_{\mathrm{b},{\vec{n}}}(\omega)=\pi\sum_ig_{{\vec{n}},i}^2\delta(\omega-\omega_{{\vec{n}},i})=\pi\lambda_{\vec{n}}\omega \exp(\omega/\omega_{\mathrm{c},{\vec{n}}})/\omega_{\mathrm{c},{\vec{n}}}$, where $\lambda$ is the reorganization energy and $\omega_\mathrm{c}$ is the cut-off frequency. While we adopt the Ohmic bath for the numerical examples presented in this contribution, the actual form of the bath spectral density is irrelevant in the proposed generalized scheme. In accordance to the homogeneous limit applied to $H_\mathrm{s}$, we set $J_{\mathrm{b},{\vec{n}}}(\omega)=J_{\mathrm{b}}(\omega)$.

The linear absorption is determined by the second order autocorrelation function of the transition dipole operator between the ground and the first excited state manifolds of the system Hamiltonian Eq.~(\ref{eqn:Hs}).\cite{mukamel}
\begin{eqnarray}
\hat{I}(t)&=&\mathrm{Tr}_\mathrm{b}[\hat{\mu}(t)\hat{\mu}(0)],I(t)=\sum_{{\vec{n}},{\vec{m}}}(\vec{\epsilon}\cdot\vec{e}_{\vec{n}})(\vec{\epsilon}\cdot\vec{e}_{\vec{m}})\langle{\vec{n}}|\hat{I}(t)|{\vec{m}}\rangle\nonumber
\end{eqnarray}
where $\hat{\mu}=\sum_n|0\rangle\langle n|+|n\rangle\langle 0|$ is expressed in the interaction picture, and $\mathrm{Tr}_\mathrm{b}$ represents tracing over the bath degrees of freedom. $\vec{\epsilon}$ is the polarization vector of the electric field of photons and $\vec{e}_{\vec{n}}$ is the transition dipole moment of molecule ${\vec{n}}$. The experimentally measured spectrum can be compared to the single-sided Fourier transform of $I(t)$, \textit{i.e.} $I(\omega)=2\mathrm{Re}[\int_0^\infty dtI(t)e^{i\omega t}]$. \cite{mukamel,LiamNJP}

Our starting point is the full second order cumulant expansion method devised previously by our group,\cite{JianFCE} which has been demonstrated to give reliable results across a wide range of system-bath coupling strength $\lambda$. The absorption operator is expressed as
\begin{eqnarray}
\hat{I}(t)&\approx& e^{-i\hat{H}_\mathrm{s}t}e^{-\hat{K}(t)}\\
\hat{K}(t)&=&\sum_{\mu,\nu}|\mu\rangle\langle\nu|\sum_\kappa\sum_{\vec{n}}X_{\vec{n}}^{\mu,\kappa}X_{\vec{n}}^{\kappa,\nu}\int_0^tdt_2\int_0^{t_2}dt_1e^{i\omega_{\mu\kappa}t_2-i\omega_{\nu\kappa}t_1}C_\mathrm{b}(t_2-t_1)\label{eqn:FCE}\\
C_\mathrm{b}(t)&=&\int_0^\infty d\omega J_\mathrm{b}(\omega)[\coth(\omega/2k_\mathrm{B}T)\cos(\omega t)-i\sin(\omega t)]
\end{eqnarray}
Here the Greek summation indices run through the system eigenstates, where $X_{\vec{n}}^{\kappa,\alpha}=|\phi_{\vec{n}}^{\kappa*}\phi_{\vec{n}}^{\alpha}|$ is the overlap between the wavefunction amplitudes $\phi_{\vec{n}}^\kappa$. $C_\mathrm{b}(t)$ is the autocorrelation function of the generalized bath coordinate, and $k_\mathrm{B}$ is the Boltzmann constant. See Ref.[\citenum{JianFCE}] for detailed descriptions. 

In order to obtain a clear physical picture and connect spectroscopic features to microscopic aggregate structures, we further make the secular and the Markovian approximations to Eq.~(\ref{eqn:FCE}). The former dictates that the lineshape operator $\hat{K}(t)$ be diagonal in the system eigenbasis, and the latter essentially extracts its long-time component that is linear in $t$. The resulting expression can be written as
\begin{eqnarray}
\hat{K}(t)&\approx&-\sum_\mu|\mu\rangle\langle\mu| \left(\Gamma_\mu^\mathrm{pd}+\sum_\kappa{}^{'}\Gamma_{\mu\kappa}^\mathrm{rd}\right)t,
\end{eqnarray}
where the pure dephasing $\Gamma_\mu^\mathrm{pd}$ and the population relaxation $\Gamma_{\mu\kappa}^\mathrm{rd}$ parts are given by
\begin{eqnarray}
\Gamma_\mu^\mathrm{pd}&=&\frac{k_\mathrm{B}T}{N}\cdot\lim_{\omega\rightarrow0}\frac{J_\mathrm{b}(\omega)}{\omega}\\
\Gamma_{\mu\kappa}^\mathrm{rd}&=&\frac{1}{N}\left\{J_\mathrm{b}(\omega_{\mu\kappa})\bar{n}_\mathrm{BE}(\omega_{\mu\kappa})+i\frac{1}{\pi}\mathcal{P}\int_0^\infty d\omega\cdot J_\mathrm{b}(\omega)\left[\frac{2\omega_{\mu\kappa}\bar{n}_\mathrm{BE}(\omega)}{\omega^2-\omega_{\mu\kappa}^2}+\frac{1}{\omega-\omega_{\mu\kappa}}\right]\right\},
\end{eqnarray}
where $\mathcal{P}$ stands for the Cauchy principal value of the integral, applied to the singularities of the integrand at $\omega=|\omega_{\mu\kappa}|$.  $\bar{n}_\mathrm{BE}(x)=(e^{x/k_\mathrm{B}T}-1)^{-1}$ is the Bose-Einstein distribution. We note that the secular approximation is automatically satisfied given a translationally invariant system, the case we assume throughout this contribution. In deriving the last two expressions we take the exciton center-of-mass wavefunction to be delocalized up to $N$ monomers within the aggregate, \textit{i.e.} $X_n^{\mu\nu}=N^{-1}$. In the homogeneous limit, the pure dephasing contribution vanishes as $N\rightarrow\infty$, and the relaxation dephasing can be cast into a more transparent form relating to the system DoS, $D_\mathrm{s}(\omega)$.
\begin{eqnarray}
\hat{K}(t)&\approx&-\sum_\mu|\mu\rangle\langle\mu|\left[W(\omega_\mu)+iS(\omega_\mu)\right]t\label{eqn:KSM}\\
W(\omega)&=&\int d\omega'\cdot D_\mathrm{s}(\omega')J_\mathrm{b}(|\omega-\omega'|)\bar{n}_\mathrm{BE}(|\omega-\omega'|)+W_0(\omega)\label{eqn:W}\\
S(\omega)&=&\frac{2}{\pi}\int d\omega'D_\mathrm{s}(\omega')(\omega-\omega')\cdot\mathcal{P}\int_0^\infty d\omega''\frac{J_\mathrm{b}(\omega'')\bar{n}_\mathrm{BE}(\omega'')}{\omega''^2-(\omega-\omega')^2}+S_0(\omega)\label{eqn:S}
\end{eqnarray}
where the $T$-independent, phonon spontaneous emission terms are expressed as
\begin{eqnarray}
W_0(\omega)&=&\int d\omega'D_\mathrm{s}(\omega')J_\mathrm{b}(\omega-\omega')\Theta(\omega-\omega')\\
S_0(\omega)&=&\frac{1}{\pi}\int d\omega'D_\mathrm{s}(\omega')\cdot\mathcal{P}\int_0^\infty d\omega''\frac{J_\mathrm{b}(\omega'')}{\omega''-(\omega-\omega')}
\end{eqnarray}
While Eq.~(\ref{eqn:W}) has been obtained by Heijs et al.\cite{Heijs2005} and utilized to explained the power law scaling of $T$-dependent absorption line width, Eq.~(\ref{eqn:S}) is the main result of the current contribution. 

A few comments are due in connecting the above formulation to experimental observables. Firstly note that owing to the diagonal (secular) form of the lineshape operator, the full spectrum can be broken down into individual bright states at frequencies $\omega_\mu$'s and their relative locations within the system DoS. Second, the Markovian approximation predicts Lorentzian lineshape for homogeneous broadening where $W(\omega)$ exactly corresponds to the full width at half maximum for a peak located at $\omega+S(\omega)$. Lastly and most importantly, while the width $W$ monotonically increases with increasing temperature, the sign of the $T$-dependent peak shift is dictated by the relative abundance of system DoS to the higher and the lower energy side of the bright state, weighted by the bath spectral density and phonon thermal occupation numbers. This is owing to the asymmetry of the dependence on energy gap $\omega-\omega'$ in Eq.~(\ref{eqn:S}). More specifically, this expression predicts that DoS higher in energy than the bright state effectively pushes it away, \textit{inducing a redshift with increasing temperature}, whereas \textit{DoS lower in energy than the bright state induces blueshift with increasing temperature}. We note that this is similar to the effects of static disorder, as is discussed in our earlier paper. \cite{ACSNano2018} In addition, this effect is additive, as the net $T$-dependent shift is determined by whichever side has a larger weighted system DoS in Eq.~(\ref{eqn:S}). This proves very useful in the generalized classification scheme proposed below and can be utilized to explain experimental results previously observed that are not accounted for by Kasha's original theory.

In the following we elaborate on the $T$-dependent lineshape analysis first for the classic Kasha 1D model. Afterward, a generic model for 2D aggregates is analyzed which shows richer behaviors in the configurational space. While Eqs.~(\ref{eqn:KSM})-(\ref{eqn:S}) connect spectroscopic features to system DoSs, one further conceptual step is needed in order to compare to Kasha's original theory of 1D aggregates: Inferring the microscopic packing conditions in 2D from the DoSs. This is presented in the following section to complete the newly proposed classification scheme before the concluding section.  

\section{Configurational Space for 1D Aggregates}
\begin{figure}[ht]
	\centering
  \includegraphics[height=11cm]{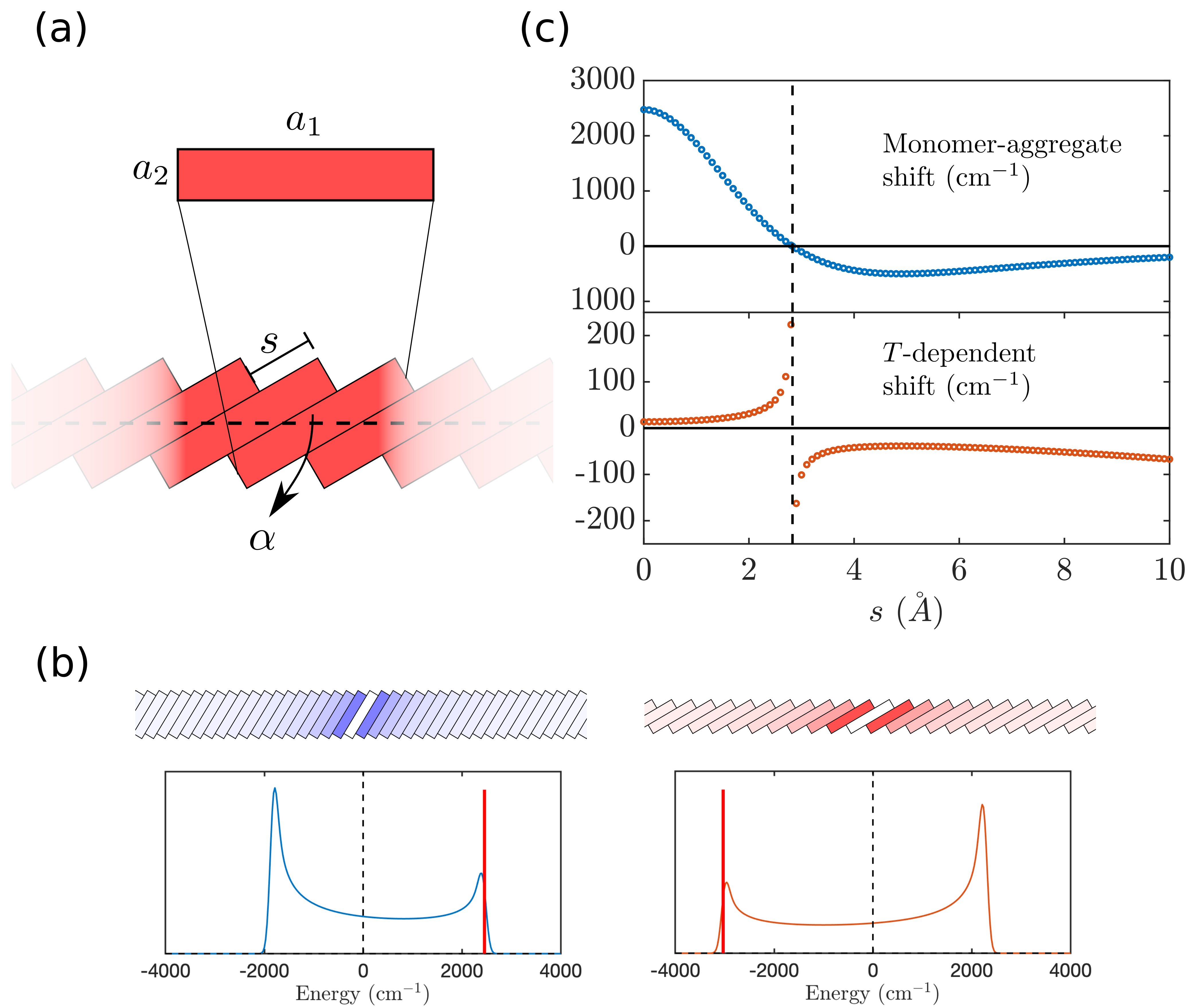}
  \caption{(a) Schematic illustration of a 1D molecular aggregate.  $s$ is the offset of adjacent molecular tiles. Here we set $(a_1,a_2)=(20,4)$ ($\AA$). See text for details. (b) (Top) Representative J- and  H-aggregates in 1D. Red- (blue-)shaded bricks correspond to negative (positive) excitonic dipole-dipole couplings with the molecule at the middle. The corresponding DoS are shown at the bottom with the vertical lines indicating the locations of the bright states. (c) (Top) Monomer-aggregate shift calculated with dipole approximation, $\mu_0=10$ (Debye). The monomer transition energy is set to zero in all cases. (Bottom right) $T$-dependent shift between $T=0$ and $T=300$ K, calculated using Eq.~(\ref{eqn:S}). We use an Ohmic spectral density $J_\mathrm{b}(\omega)=\pi\lambda\frac{\omega}{\omega_\mathrm{c}}e^{-\omega/\omega_\mathrm{c}}$, with reorganization energy $\lambda=1000$ and cut-off frequency $\omega_\mathrm{c}=1000$ in wavenumber.}
  \label{fig:1D}
\end{figure}

In this section we apply the $T$-dependent spectral lineshape analysis to Kasha's classic 1D model for illustrative purposes. As shown in Fig.\ref{fig:1D}(a) we represent a simplified 1D aggregate by a row of slanted rectangular bricks, which correspond to the area occupied by the molecules projected onto the plane of aggregation. Assuming a large aspect ratio of the bricks ($a_1/a_2\gg1$) and transition dipoles parallel to the long axes, this model is equivalent to the original Kasha model, while it can also be intuitively extended to 2D aggregates as demonstrated in the next section. Given a slip parameter $s$, an offset in the direction of the molecular long axis between nearest neighbors, a unique configuration can be specified with $\alpha=90^\circ-\arctan(s/a_2)$, \textit{i.e.} the angle between the molecular long axis and the direction of aggregation, and $x_0=a_2\csc\alpha$, \textit{i.e.} the lattice spacing. Fig.~\ref{fig:1D}(b) shows typical 1D J- and H-aggregates with their corresponding DoS. 

In Fig.~\ref{fig:1D}(c) we present the dependences of both the monomer-aggregate peak shift and the $T$-dependent shift, given in Eq.~(\ref{eqn:S}), on the slip parameter $s$. The former (upper panel) follows Kasha's rule, where the monomer-aggregate shift for 1D aggregates can be written as $\Delta E(\alpha)=2J_\mathrm{nn}(\alpha)\xi(3)$, where $J_\mathrm{nn}(\alpha)=C\mu_0^2(1-3\cos^2\alpha)/x_0^3$ is the nearest neighbor coupling strength and $\xi(3)\approx1.202$ is the Riemann zeta function. $\mu_0$ is the magnitude of the transition dipole in Debye and $C\approx5040$ is the conversion factor taking $x_0$ in Angstr\"{o}m and $J_\mathrm{nn}$ in wavenumber. Given the simple dipole form of the excitonic interactions, the excitonic shift is a monotonic function of $s$, specifically positive shifts (H-aggregates) for smaller $s$ and negative shifts (J-aggregates) for larger ones. The transition occurs at $\alpha=\theta_\mathrm{m}=\arccos(1/\sqrt{3})\approx54.7^\circ$, the magic angle defined in the context of spectroscopies. 

The sign of the $T$-dependent shift, shown in the lower panel of Fig.~\ref{fig:1D}(c), follows a trend identical to that of the excitonic shift. In other words, 1D J-aggregates that are redshifted compared to the monomers undergo \textit{an additional redshift upon increasing temperature}, while the opposite is true for 1D H-aggregates. This follows from our earlier discussion on the relative position of the bright state ($k=0$ nodeless Bloch wave) in the system DoS: The bright state of a 1D J- (H-)aggregate is located at the bottom (top) of the exciton band under the dipole approximation, as shown in Fig.~\ref{fig:1D}(b). Thus, in Fig.~\ref{fig:1D}(c), both the monomer-aggregate (top panel) and the $T$-dependent shift change sign at the same numerical value of $s$, which corresponds to the magic angle of $\alpha=\theta_\mathrm{m}$. In other words, the $T$-dependent shift fully correlates with the monomer-aggregate excitonic shift in 1D aggregates and \textit{does not provide additional qualitative information with regard to the microscopic packing conditions.} We shall see how this simple picture is changed in a generic 2D molecular aggregate in the next section.

\section{Configurational Space for 2D Aggregates}
\begin{figure}[ht]
	\centering
  \includegraphics[height=12cm]{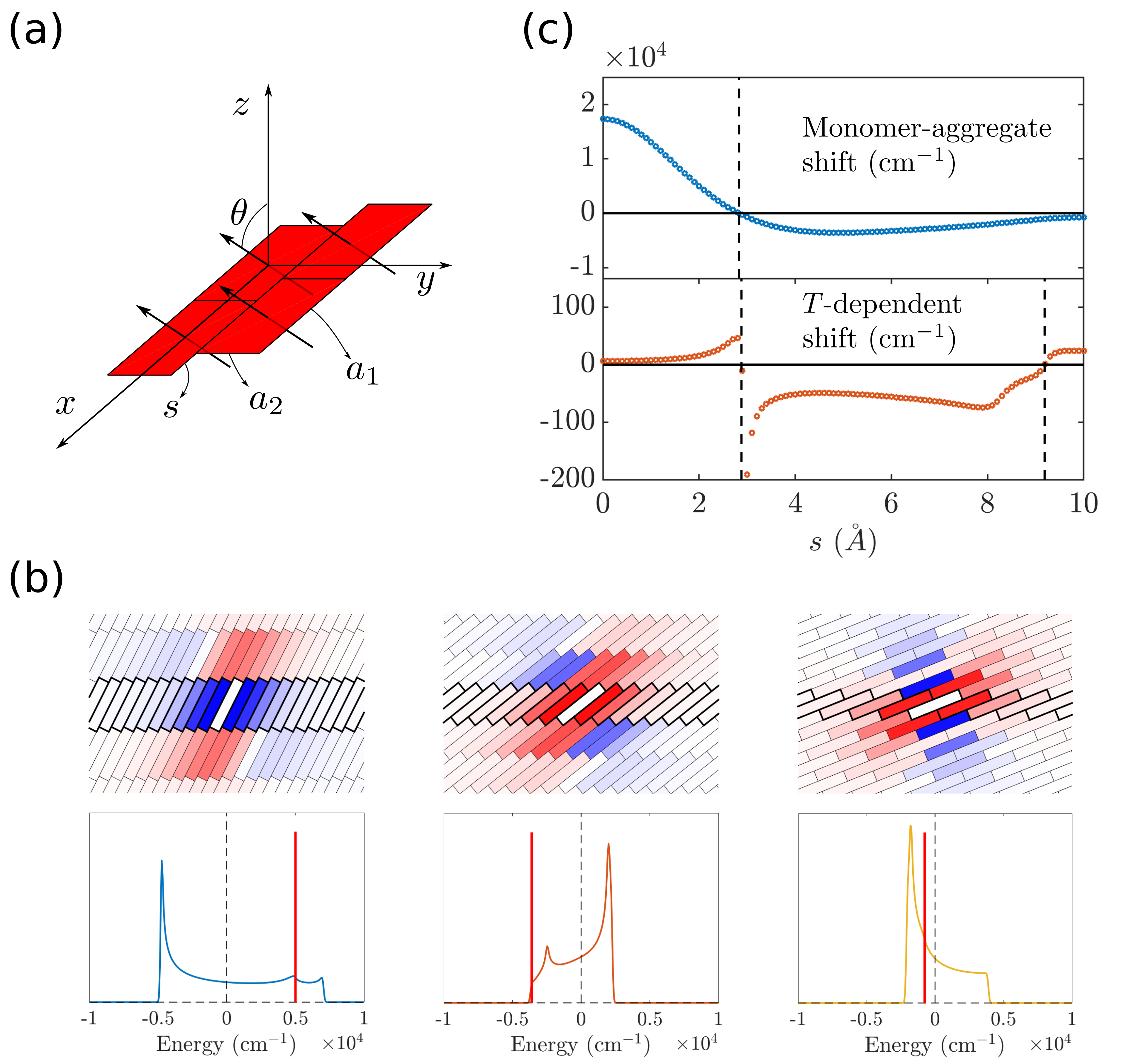}
  \caption{(a) Illustration of the 2D brick-layer geometry and the coordinate system. (b) (Top) Representative 2D aggregates with different slip parameters: $s=2,~5,~10~\AA$ from left to right. Red- (blue-)shaded bricks correspond to negative (positive) excitonic dipole-dipole couplings with the molecule at origin. The 1D aggregates with the same slip parameters are outlined with thick frames. (Bottom) The DoS of the three aggregates drawn on the top. The locations of the bright states $\vec{k}=0$ are marked by the vertical red lines. (c) The two-panel configurational space diagram for the 2D aggregates in question. All the conventions follow those in Fig.~\ref{fig:1D}(c). In all of the aggregates we adopt the same molecular dimension $(a_1,a_2)=(20,4)~\AA$ and set the transition dipole moment making an angle of $\theta=70^\circ$ with the $z$-axis.}
  \label{fig:2D}
\end{figure}

By directly extending the corresponding 1D aggregates to the 2D plane with various slip parameters under the close-packing condition, one can define 2D aggregates as shown in Fig.~\ref{fig:2D}(a). A defining character of an excitonic 2D molecular aggregate is the \textit{incommensurability between the anisotropic in-plane dipolar interactions and the 2D geometry}.\cite{Spano2012,LiamPNAS} In other words, with the exception of the case with all transition dipoles are pointing out of the plane of aggregation ($\theta=90^\circ$) and all couplings are positive,\cite{Rasmussen1998,Malyshev2D}, \textit{i.e.} a "perfect" H-aggregate, positive and negative excitonic couplings coexist. More specifically, any molecule with a position vector making an angle $|\alpha|<\theta_\mathrm{m}$ with the direction of the transition dipoles couple negatively with a test molecule at the origin and vice versa. 

In addition, the discreteness of the lattice gives rise to a plethora of plausible DoS with variable bright state locations, owing to the sensitive dependence of the short-range couplings on the slip parameter. The DoS and the bright state locations corresponding to the lattices shown are presented on the right hand side of Fig.~\ref{fig:2D}(b). Note that by definition aggregates with bright states located higher in energy than zero (monomer transition energy) are considered H-aggregates and vice versa. At the bottom we have an H-aggregate ($s=2~\AA$), while the upper two ($s=5,~10~\AA$) are J-aggregates.

Our first observation is that the bright state is not necessarily located at either end of the DoS, in stark contrast to the 1D aggregates as shown in Fig.~\ref{fig:1D}(b). Importantly, this opens up the possibility of a different trend for the $T$-dependent shift, Eq.~(\ref{eqn:S}), with respect to the excitonic shift. Indeed, as demonstrated in Fig.~\ref{fig:2D}(c), the same two-panel configurational space diagram used in Fig.~\ref{fig:1D}(c) for the family of 2D aggregates shown in Fig.~\ref{fig:2D}(b). Notice that the small $s$ regions of Fig.\ref{fig:1D}(c) and Fig.\ref{fig:2D}(c) are rather similar due to the large aspect ratio $a_1/a_2$ and the close-packing condition assumed, where the couplings in the horizontal direction (shown in thick frames in Fig.~\ref{fig:2D}(b)) dominate in determining the DoS. However, a novel regime appears at $s>9.2~\AA$ where J-aggregates exist that blueshift upon increasing temperature. J-aggregates with this peculiar property have been reported experimentally without explanation.\cite{Yamaguchi2006,Wurthner2009} In fact, the aggregate shown on the right of Fig.~\ref{fig:2D}(b) is one such example. In accordance with our analysis above, the existence of a blueshifting J-aggregate (BJ-aggregate) requires larger DoS to the lower energy side of the bright state, as seen in the bottom right of Fig.~\ref{fig:2D}(b).

\section{Short and Long-Range Interactions in 2D Aggregates}
\begin{figure}[ht]
	\centering
  \includegraphics[height=8cm]{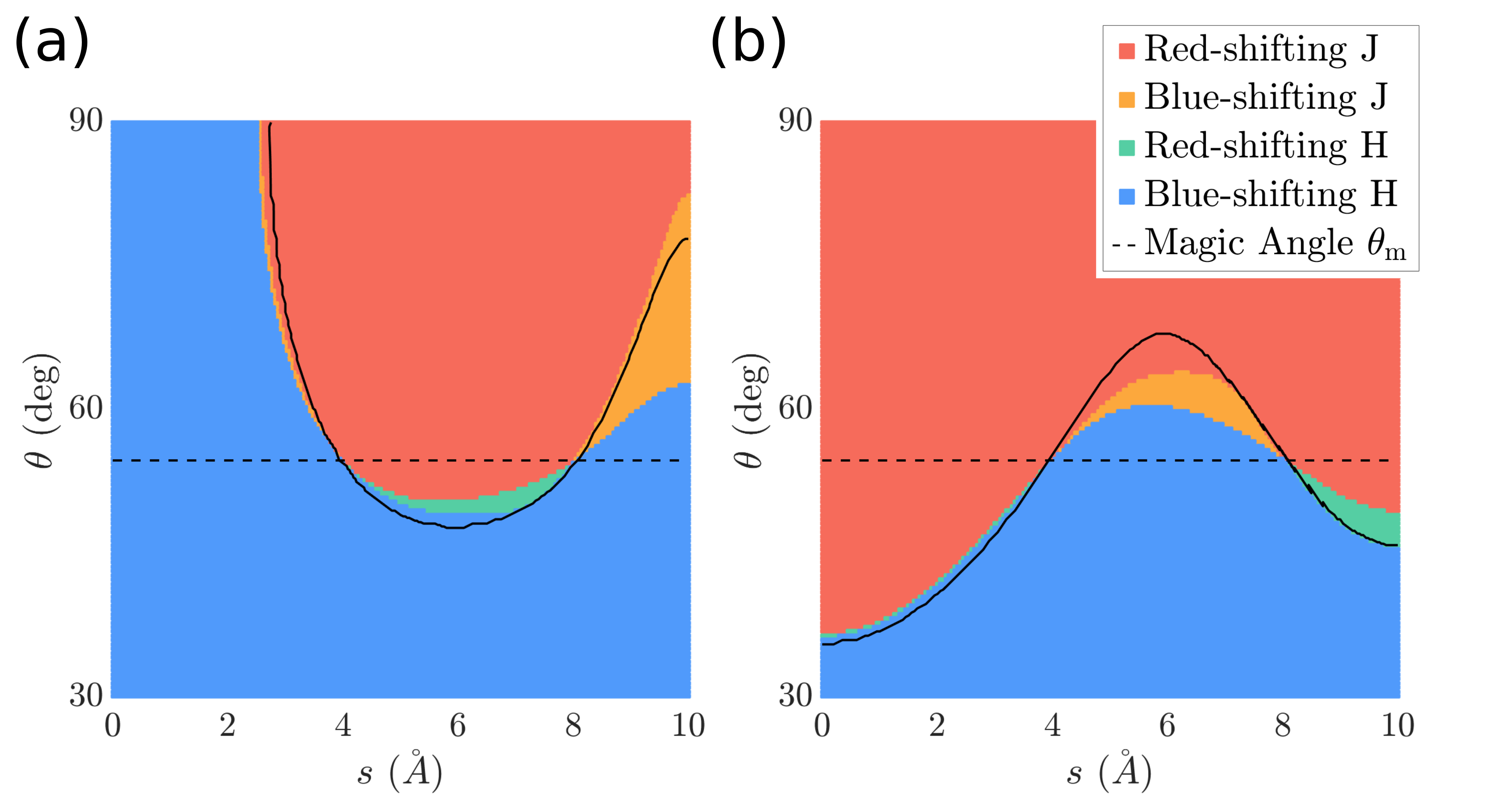}
  \caption{(Color) 2D configurational space diagrams of the four possible combinations of monomer-aggregate and $T$-dependent shifts. The molecular dimension is the same as in previous figures. The horizontal axis is the slip parameter $s$ and the vertical axis is the angle $\theta$ between the transition dipole and the $z$-axis protruding out-of-plane of the aggregate. (a) Transition dipole lying along the long axis ($a_1$) of the molecular frame. (b) Transition dipole lying along the short axis ($a_2$). The solid lines indicates the geometries corresponding to vanishing net short-range couplings. We omit the region $0<\theta<30^\circ$ since it is entirely occupied by BH-aggregates.}
  \label{fig:2Dpd}
\end{figure}

An immediate question arises as to whether or not a redshifting H-aggregate (RH-aggregate) exists. We give a positive answer by further exploring the configurational space under the current scheme of close-packing rectangles with two parameters: the slip $s$ and the zenith angle $\theta$. Here $\theta=0$ corresponds to the perfect H-aggregate with all dipoles pointing out of the plane and $\theta=90^\circ$ pointing completely in-plane. All four combinations of monomer-aggregate shift and $T$-dependent shift with increasing temperature are recorded: RJ-, BJ-, RH-, and BH-aggregates as shown in Fig.~\ref{fig:2Dpd}. In Fig.~\ref{fig:2Dpd}(a) the in-plane component of the transition dipoles is again set to be parallel to the long axes of the molecules as in most cyanine dyes, while in Fig.~\ref{fig:2Dpd}(b) the dipoles are parallel to the short axes, exemplified by oligoacenes.

Several notes are in order in interpreting the 2D configurational space. First, the general trend of monomer-aggregate shift (J- or H-) along the vertical axis follows intuitively: The larger the out-of-plane component (small $\theta$) the larger the blueshift. Closer to the magic angle, the dependence on the slip parameter becomes more significant. This can be understood by separating the long- and short-range contributions to the bright state energy ($E_\mathrm{b}$):
\begin{eqnarray}
E_\mathrm{b}&=&E_0+\sum_{\vec{n}}J_{\vec{n}}=E_0+\left(\sum_{\vec{n}\in\mathrm{s.r.}}+\sum_{\vec{n}\in\mathrm{l.r.}}\right)J_{\vec{n}}\nonumber\\
&\approx&E_0+\sum_{\vec{n}\in\mathrm{s.r.}}J_{\vec{n}}+A\rho_0^2\int_a^{\infty}dr\int_0^{2\pi}r~d\phi\cdot\frac{1}{r^3}\left(1-3\sin^2\theta\cos^2\phi\right)\nonumber\\
&=&E_0+E_\mathrm{b}^\mathrm{s.r.}+\frac{\pi\rho_0^2}{a}\left(2-3\sin^2\theta\right)\label{eqn:Eb}
\end{eqnarray}
where $E_0$ is the bare transition energy including the solution-to-aggregate shift. In the second line we treat the long-range contribution with the continuum approximation and a transition dipole area density $\rho_0=\mu_0/A$ is defined, where $A=a_1a_2$ is the area of the molecule occupied in the plane of aggregation. We also set a cut-off radius $a$ comparable to the dimension of the molecule and an in-plane component parallel to the $x$-axis. In practice the choice of the cut-off radius $a$ depends on the nature of the excitonic coupling, as will be discussed in the next section. It is clear that the long-range contribution follows exactly the above mentioned trend and is independent of the slip parameter, owing to the continuum approximation made. The deviation of the curves separating the J- and the H-aggregates in Fig.~\ref{fig:2Dpd} from $\theta=\theta_\mathrm{m}$ is due to the sensitive dependence of $E_\mathrm{b}^\mathrm{s.r.}$ on $s$.

In contrast, the two unconventional types split perfectly on either side of the line. Specifically, BJ-aggregates exist only for $\theta>\theta_\mathrm{m}$ and RH-aggregates exist only for $\theta<\theta_\mathrm{m}$. 
We again invoke the separation of long- and short-range interactions to explain this observation. Since the $T$-dependent shift is essentially the result of dephasing by elastic scattering of phonons between excitonic states, we require not only information on the bright state but also information on the system DoS near the bright state. The exciton dispersion relation is given by
\begin{eqnarray}
E_{\vec{k}}&=&E_0+\sum_{\vec{n}}J_{\vec{n}}\cos\left(\vec{k}\cdot\vec{n}\right)=E_0+E_{\vec{k}}^\mathrm{s.r.}+E_{\vec{k}}^\mathrm{l.r.},\label{eqn:Ek}
\end{eqnarray}
where
\begin{eqnarray}
E_{\vec{k}}^\mathrm{s.r.}&=&\sum_{\vec{n}\in\mathrm{s.r.}}J_{\vec{n}}\cos\left(\vec{k}\cdot\vec{n}\right)\label{eqn:Eksr}\\
E_{\vec{k}}^\mathrm{l.r.}&=&\sum_{\vec{n}\in\mathrm{l.r.}}J_{\vec{n}}\cos\left(\vec{k}\cdot\vec{n}\right)\approx A\rho_0^2\int~d\vec{r}~e^{i\vec{k}\cdot\vec{r}}\frac{1-3\sin^2\theta\cos^2\phi}{r^3}\label{eqn:Eklr}.
\end{eqnarray}
To determine the direction of the $T$-dependent spectral shift, we need to connect Eq.~(\ref{eqn:Ek}) to the system DoS as indicated by Eq.~(\ref{eqn:S}). This is achieved with key observations about $E_{\vec{k}}^\mathrm{s.r.}$ and $E_{\vec{k}}^\mathrm{l.r.}$. For the former, as in Eq.~(\ref{eqn:Eksr}), the cosine form implies that the existence of van Hove singularities, points in the DoS that are not differentiable, near the bright state $\vec{k}=0$. These singularities are attenuated by the long-range coupling contribution $E_{\vec{k}}^\mathrm{l.r.}$, since the band structure of 2D dipole-dipole interaction in the continuum is known to be linearly dispersed near $\vec{k}=0$,\cite{Rasmussen1998,Malyshev2D} leading to a smooth and slowly varying contribution to the DoS.

Owing to this generic scaling property of $E_{\vec{k}}^\mathrm{l.r.}$, we deduce that the relative abundance of DoS on either side of the bright state and, therefore, the direction of the $T$-dependent shifting, is dominated by the short-range contribution to the band structure $E_{\vec{k}}^\mathrm{s.r.}$. In other words, \textit{redshifting aggregates are those with net negative short-range couplings and blueshifting ones are with net positive short-range couplings.} Moreover, this is so regardless of the overall net coupling that defines J- and H- aggregates. To further justify this argument, we calculate the following quantity that serves as a measure of net short-range coupling:
\begin{eqnarray}
J^\mathrm{s.r.}=\sum_{\vec{n}}M(\vec{r}_{\vec{n}})J_{\vec{n}}.
\label{eqn:Jsr}
\end{eqnarray}
Notice that the excitonic shift, $E_\mathrm{b}-E_0$, is given by the same expression with $M=1$. In general the function $M(\vec{r})$ should be bounded within the interval $[0,1]$ and monotonically decreasing with $|\vec{r}|$, since the contribution with large $|\vec{r}|$ is well approximated by Eq.~(\ref{eqn:Eklr}). The choice of $M(\vec{r})$ as a smooth function is to avoid discrete jumps in the solution to $J^\mathrm{s.r.}(s,\theta)=0$. Also, $M(\vec{r})$ depends on the nature of the coupling matrix element $J_{\vec{r}}$, specifically the transition charge distributions of the dyes. For the simple dipole interaction used in Fig.~\ref{fig:2Dpd}, we find $M(\vec{r})=\arctan(-|\vec{r}|/a)$ with $a=1~\AA$ to be a valid working definition. Shown as the solid lines in Fig.~\ref{fig:2Dpd} are the solutions to the equation $J^\mathrm{s.r.}(s,\theta)=0$. Apart from the deviations at large $|\theta-\theta_\mathrm{m}|$, Eq.~(\ref{eqn:Jsr}) correlates quantitatively with the $T$-dependent shift, Eq.~(\ref{eqn:S}). This provides a direct connection from the packing geometry to spectroscopic properties. Finally, we note that the limited configurational space volumes occupied by BJ- and RH-aggregates is likely due to the simple dipole approximation assumed in Fig.~\ref{fig:2Dpd}. In our experience going beyond the dipole approximation often leads to more significant configurational space volumes for the two unconventional aggregates, as exemplified by the case study detailed in the next section.


\section{Spectroscopic Inference of Microscopic Packing Geometry}
We are now at a position to explain the above mentioned characteristics of the 2D configurational space diagrams shown in Fig.~\ref{fig:2Dpd} and to connect to the microscopic packing conditions. Starting from the two conventional types, the RJ- and the BH-aggregates, each of them could be the result of two possible combinations of long- and short-range couplings. Firstly, both the long and the short-range contributions are of the same sign. This accounts for the case of RJ- (BH-) aggregate for $\theta>\theta_\mathrm{m}$ ($\theta<\theta_\mathrm{m}$). On the other hand, RJ- (BH-) aggregates could also exist when the two contributions are of opposite signs. Specifically, this happens when the dominant contribution to the excitonic shift is the short-range part. For example, an RJ-aggregate could result from a lattice structure and transition charge density distribution supporting strongly negative short-range coupling even when $\theta<\theta_\mathrm{m}$, \textit{i.e.} positive long-range coupling. On the one hand, it is a J-aggregate because of the net negative coupling combining the two contributions. On the other hand, it is redshifting due to the short-range contribution being negative. The same argument applies to BH-aggregates with $\theta>\theta_\mathrm{m}$.


As for the two unconventional types, the BJ- and RH-aggregates, the situation is reversed. While it is clear that the short and the long-range parts of excitonic coupling are of different signs, the latter is the dominant contribution. However, since the $T$-dependent shift is dictated by the short-range couplings, this leads to the said novel $T$-dependent properties of the corresponding aggregates and the strict separation between the BJ- and the RH-aggregate regions in configurational space by the line $\theta=\theta_\mathrm{m}$. We summarize the above discussion in Table.\ref{tab:2Dtab}. 

\begin{table}[ht]
\def\arraystretch{1.2}
\begin{tabular}{|c|c|c|c|}
\hline
Type			& Long-range	& Short-range	& Dominant coupling	\\ \hline
RJ			& J/H		& J			& Short-range		\\ \hline
BJ			& J			& H			& Long-range		\\ \hline
RH			& H			& J			& Long-range		\\ \hline
BH			& J/H		& H			& Short-range		\\ \hline
\end{tabular}
  \caption{All possible combinations of the signs of the long- and the short-range couplings in determining the monomer-aggregate and the $T$-dependent shifts. Note that the sign of the long-range coupling is entirely determined by $\theta$. Specifically long-range negative couplings correspond to $\theta>\theta_\mathrm{m}$ and vice versa.}
  \label{tab:2Dtab}
\end{table}

\begin{figure}[ht]
	\centering
  \includegraphics[height=6cm]{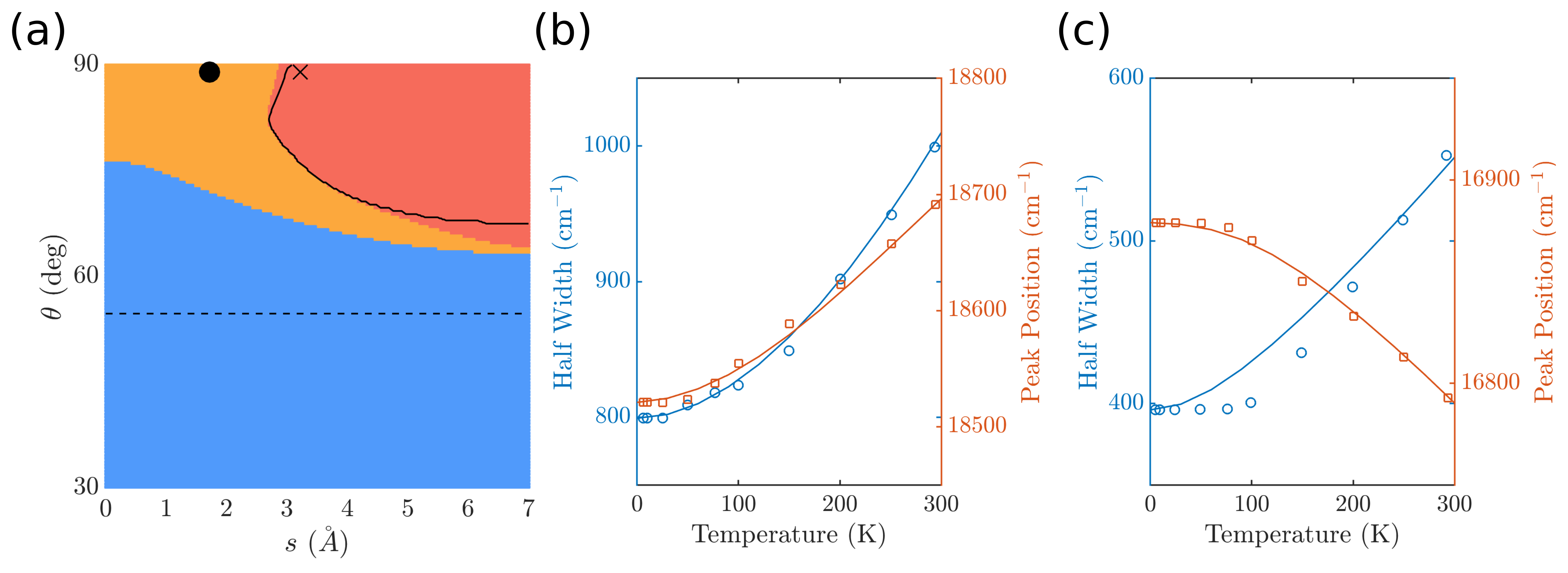}
  \caption{(a) 2D configurational space diagram for the asymmetric merocyanine dyes MC(O) monolayers studied in Ref.[\citenum{Yamaguchi2006}]. The excitonic coupling matrix element $J_{\vec{n}}$ is calculated using the transition monopole method\cite{ChunTeh2014}, and the lattice constants $(a_1,a_2)=(14,5)~\AA$. (b) The $T$-dependent absorption width (blue) and shift (orange) of the MC(O) 2D aggregate. The symbols are data taken from experiments and lines are the results of Eqs.~(\ref{eqn:W}) and (\ref{eqn:S}). The lattice parameters are $(s,\theta)=(1.7~\AA,90^\circ)$, marked as the round dot in (a). The phonon bath is described by an Ohmic spectral density with $\omega_\mathrm{c}=1000~\mathrm{cm}^{-1}$ and $\lambda=3700~\mathrm{cm}^{-1}$. An inhomogeneous broadening of magnitude 110 cm$^{-1}$ is added to the absorption width. (c) The $T$-dependent spectral trends for the MC(S) 2D aggregate, with lattice parameters $(s,\theta)=(3.2~\AA,90^\circ)$, marked as the cross in (a). An Ohmic bath with $\omega_\mathrm{c}=1000~\mathrm{cm}^{-1}$ and $\lambda=1400~\mathrm{cm}^{-1}$ and an inhomogeneous width 350 cm$^{-1}$ are used.}
  \label{fig:Kometani}
\end{figure}

With the classification framework set, it is helpful to apply the theory to real systems. In Fig.\ref{fig:Kometani} we study the $T$-dependent absorption peak width and shift of asymmetric merocyanine dye monolayers that have been characterized experimentally\cite{Yamaguchi2006}. Three different dyes with similar structures are considered: MC(X) (X=O, S, Se). While all three dyes form J-aggregate monolayers, it is most noteworthy that the MC(O) aggregate shows a different $T$-dependent trend than the other two. Specifically the MC(O) aggregate absorption peak blueshifts with increasing temperature, an example of 2D BJ-aggregate. 

In Fig.\ref{fig:Kometani}(a) we show the 2D configurational space diagram of the MC(O) dye 2D aggregate. While the lattice structure is similar to that adopted in Fig.\ref{fig:2Dpd}(a) with the transition dipole parallel to the molecular long axis, we find no configuration corresponding to an RH-aggregate. This is likely due to the nature of the transition charge distribution of the merocyanine dye molecules, where we find quantitatively similar transition charge distributions for MC(S) and MC(Se) dyes as well (data not shown). Thus, the solid line corresponding to $J^\mathrm{s.r.}(s,\theta)=0$ nearly overlaps with the boundary of RJ-aggregate configurational space. In Fig.\ref{fig:Kometani}(b) and (c) we show the comparisons between the experimental $T$-dependent lineshapes of MC(O) and MC(S) aggregates and those from Eqs.~(\ref{eqn:W}) and (\ref{eqn:S}). In both cases we set $\theta=90^\circ$, while the corresponding slip parameters marked in (a). While there are quantitative agreements between experiment and theory in general, we find noticeable deviation of low temperature absorption width in the MC(S) case. This is attributed to the \textit{ad hoc} treatment of sizable static disorder and will be addressed in our future work.

\section{Discussion}
To sum up, we demonstrate that the interplay between the relative signs and strengths of short and long-range couplings is the key factor in understanding the expanded classification scheme. We emphasize that the long-range couplings are qualitatively determined by a single parameter: the zenith angle $\theta$ of the transition dipole moment. This parameter can be assessed robustly by the chemical constitution of the monomers. For example, for the most studied family of excitonic molecular aggregates, amphiphilic cyanine dyes,\cite{Kirstein2006} the hydrophilic and the hydrophobic functional groups are at the sides of the conjugated backbone. This unambiguously defines the plane of aggregation, accommodating a fully in-plane transition dipole moment, \textit{i.e.} $\theta=90^\circ$. Other examples allowing access to the value of $\theta$ include polarization-resolved spectroscopy of Langmuir-Blodgett films\cite{Witten1994,Nabetani1995,Yamaguchi2006} and near-field scanning optical microscopy of PIC aggregates.\cite{Barbara1996} With this information, one can deduce the dominance of either the positive or the negative short-range couplings and narrow down the possible in-plane packing geometries of the molecules. For example, blueshifting aggregates are likely the results of significant face-to-face stacking among near-neighboring dyes and vice versa.

Before concluding, we clarify related concepts previously proposed in the literature and a few complications when applying the current scheme to real systems. Spano et al. proposed an HJ-aggregate model to understand the photophysics of semiconducting polymer assemblies.\cite{Spano2012} They study the interplay between the through-bond interactions along the polymer backbone (analogous the negative, J-aggregate-like excitonic couplings in one direction) and the through-space interactions between different chains (analogous to the positive, H-aggregate-like excitonic couplings in the perpendicular direction). In addition, they have documented the existence of CT-mediated couplings for certain classes of molecular dyes that can be a critical factor in determining the monomer-aggregate shift.\cite{SpanoACR2017,SpanoCR2017} In this case, the competition of positive and negative couplings does not originate from different directions in an aggregate but from the different scaling properties of the dipole and the CT-mediated coupling components. This allows the existence of HJ- (JH-) aggregates even in 1D. 

We argue that both of these cases can be incorporated under our framework. In the former one works with a 2D aggregate model with $(s,\theta)=(0,90^\circ)$ and suitably includes the exchange (through-bond) interaction in the $x$-direction, shown in Fig.~\ref{fig:2D}(a). In the latter one takes into account an addition CT-mediated coupling term in a 1D dipole-coupled excitonic aggregate. Owing to the different dependence on the slip parameter $s$ of the two terms, we expect the existence of either RH- or BJ-aggregates even for 1D models. In either case, since we make no distinction of the origin of the short-range couplings, Eqs.~(\ref{eqn:Eb}) and (\ref{eqn:Eksr}) can be straightforwardly applied to the through-bond interaction in polymers and the CT-mediated interactions in the relevant dye aggregates. In fact, the nature of the short-range couplings can be distinguished into two types: The deviation from simple dipole-dipole interactions and the breakdown of the continuum approximation in describing discrete lattices. While the through-bond and the CT-mediated couplings are classified into the former, the latter is equally critical in unraveling the packing conditions. For instance, for molecules with large in-plane aspect ratios, the next-nearest-neighbor couplings in the direction of the short axis may be more dominant than the nearest-neighbor ones along the long axis. 


Lastly, we note that it is commonly found that the aggregate in question hosts two molecules per translational unit cell. 
The complexity of performing the above analysis scales quadratically with the number of transitions involved per unit cell. However, the generic consideration of separating long and short-range couplings remains a useful technique in helping to determine packing conditions. In essence, the long-range parts of Eqs.~(\ref{eqn:Eb}) and (\ref{eqn:Ek}) can be generalized to cases with $N$ transitions per unit cell, by explicitly constructing an $N$-by-$N$ coupling matrix in the $k$-space. A follow up paper will explore this possibility.

\section{Conclusion}
We report a generalized classification scheme for excitonic molecular aggregates. By expanding the original scheme proposed by Kasha to include the $T$-dependent spectral shift using standard spectral lineshape theory, we explain the origin of a class of J-aggregates that blueshift with increasing temperature, BJ-aggregates, that has been observed in the experimental literature. In addition, another novel class of redshifting H-aggregates (RH-aggregates), not yet reported to the best of our knowledge, is predicted together with configurations that might produce such cases. The $T$-dependent shift is directly related to the relative location of the bright state in the system DoS, which is a robust characteristic of the delocalized Frenkel exciton band structure, in addition to the monomer-aggregate peak shift. The usefulness of the generalized scheme is most apparent when considering the direction of the $T$-dependent shift as a binary indicator, with the magnitude of the shift quantified by Eq.~(\ref{eqn:S}).

We also establish the connection between the DoS and the microscopic packing conditions, given the knowledge of both the monomer-aggregate and the $T$-dependent shifts. This is achieved by separating the contributions to the exciton band structure into long and short-range coupling terms. The long-range contribution provides a smooth background for the DoS and is dictated by a single parameter $\theta$, the angle between the transition dipole moment and the $z$-axis. On the other hand, the short-range contribution is the dominant factor in determining the direction of the $T$-dependent shift. Specifically, a net negative short-range coupling gives rise to a redshifting aggregate. On the other hand, a net positive total short-range coupling leads to a blueshifting aggregate. Detailed analysis together with the knowledge of $\theta$ further distinguish the conventional RJ- and BH-aggregates into four classes depending on synergy between the short- and the long-range couplings. 

The generalized Kasha's scheme uses experimentally accessible information including $T$-dependent linear absorption and polarization-resolved spectroscopy, and reveals microscopic packing conditions of 2D aggregates as in Kasha's original theory of 1D aggregates. Computational information can also be seamlessly integrated in, such as the inclusion of non-dipolar interactions or even CT-mixed contributions that can greatly alter the short-range coupling but decouple from the long-range part. Our work should be of great use in deciphering aggregate geometries that evade traditional high-resolution structure-determination techniques such as X-ray scattering, and may provide a set of design principles for excitonic aggregates with desired spectroscopic properties.

\bibliography{main.bib}


\end{document}